\documentclass[aps,pra,twocolumn,nofootinbib,longbibliography]{revtex4-2}
\pdfoutput=1

\usepackage{epsf}
\usepackage{subfigure}
\usepackage[colorlinks, linkcolor =blue]{hyperref} 
\usepackage{amsmath}
\usepackage{amssymb}
\usepackage{graphicx}
\usepackage{epstopdf}
\usepackage{color}
\usepackage{enumerate}
\usepackage{amsthm} 
\usepackage{empheq}
\usepackage{array}

\newcommand{\ket}[1]{\left\lvert #1 \right\rangle}

\newcommand{\bol}[1]{\boldsymbol{#1}}
\newcommand{\Real}{\operatorname{Re}}
\newcommand{\Imag}{\operatorname{Im}}

\newcolumntype{M}[1]{>{\centering\arraybackslash}m{#1}}
\newcolumntype{N}{@{}m{0pt}@{}}

\begin{document}


\title{Unbounded quantum backflow in two dimensions}

\author{Maximilien Barbier$^{1,2}$, Arseni Goussev$^{3}$, and Shashi C. L. Srivastava$^{4,5}$}

\affiliation{$^{1}$Scottish Universities Physics Alliance, Institute of Thin Films, Sensors and Imaging, University of the
West of Scotland, Paisley PA1 2BE, Scotland, United Kingdom \\
$^{2}$Max-Planck-Institut f\"ur Physik komplexer Systeme, N\"othnitzer Str. 38, D-01187 Dresden, Germany \\
$^{3}$School of Mathematics and Physics, University of Portsmouth, Portsmouth PO1 3HF, United Kingdom \\
$^{4}$Variable Energy Cyclotron Centre, 1/AF Bidhannagar, Kolkata 700064, India \\
$^{5}$Homi Bhabha National Institute, Training School Complex, Anushaktinagar, Mumbai - 400094, India}


\begin{abstract}
Quantum backflow refers to the counterintuitive fact that the probability can flow in the direction opposite to the momentum of a quantum particle. This phenomenon has been seen to be small and fragile for one-dimensional systems, in which the maximal amount of backflow has been found to be bounded. Quantum backflow exhibits dramatically different features in two-dimensional systems that, contrary to the one-dimensional case, allow for degenerate energy eigenstates. Here we investigate the case of a charged particle that is confined to move on a finite disk punctured at the center and that is pierced through the center, and normally to the disk, by a magnetic flux line. We demonstrate that quantum backflow can be unbounded (in a certain sense), which makes this system a promising physical platform regarding the yet-to-be-performed experimental observation of this fundamental quantum phenomenon.
\end{abstract}

\noindent 
\vskip 0.5 cm

\maketitle


\section{Introduction}\label{AA}

The principle of quantum superposition is at the heart of quantum theory. Some of its best known manifestations include the double-slit interference, Schr\"odinger's cat states, and entanglement \cite{ALPP22}. Quantum backflow (QB) is another, far less known manifestation of the superposition principle. The gist of the QB effect is the counter-intuitive possibility for a quantum particle to move, in a certain sense, in the direction opposite to its momentum.

Originally, the QB problem was formulated for a free nonrelativistic particle on a line \cite{BM94}. The formulation proceeds as follows. A particle of mass $\mu$ moves freely along the $x$ axis, and its time-dependent wave function $\psi(x,t)$ is comprised of only positive-momentum plane waves:
\begin{equation}
	\psi(x,t) = \frac{1}{\sqrt{2 \pi \hbar}} \int_0^{\infty} dp \, \widetilde{\psi}(p) e^{i p x / \hbar - i p^2 t / 2 \hbar \mu} \,,
	\label{positive-momentum_psi_line}
\end{equation}
where $\widetilde{\psi}$ is the initial ($t=0$) momentum-space wave function of the particle. Eq.~\eqref{positive-momentum_psi_line} guarantees that the outcome of any momentum measurement performed on $\psi$ is bound to be positive. The wave function is normalized to unity,
\begin{equation*}
	\int_{-\infty}^{\infty} dx \, |\psi(x)|^2 = \int_{0}^{\infty} dp \, |\widetilde{\psi}(p)|^2 = 1 \,.
\end{equation*}
Surprisingly, even though the momentum of the particle is (with certainty) positive, the corresponding probability current $j(x,t)$, given by
\begin{equation}
	j(x,t) = \frac{\hbar}{\mu} \Imag \left( \psi^* \frac{\partial \psi}{\partial x} \right) \,,
\label{j_line}
\end{equation}
can be negative at some $x$ and $t$: this fact is the essence of the QB effect. A natural way to quantifying the strength of the effect is to consider the total probability transfer $\Delta$ through a fixed spatial point, say $x=0$, over a fixed (but arbitrary) time interval, say $-T/2 < t < T/2$:
\begin{equation}
	\Delta = \int_{-T/2}^{T/2} dt \, j(0,t) \,.
\label{Delta}
\end{equation}
Numerical investigations have shown~\cite{BM94,EFV05,PGK06} that $\Delta$ admits a lower bound, namely
\begin{equation}
	\inf_{\psi} \Delta = \Delta_{\text{line}} \simeq - 0.0384517 \,,
\label{Delta_line}
\end{equation}
commonly referred to as the Bracken-Melloy bound. As of today, the exact value of $\Delta_{\text{line}}$ remains unknown (lower and upper bounds for $\Delta_{\text{line}}$ have been obtained very recently~\cite{TLN22}).

Eq.~\eqref{Delta_line} shows that the effect of QB is relatively weak for a particle on a line: only a tiny fraction, less than $4\%$, of the total probability can potentially be transported in the ``wrong'' direction. This limitation is exacerbated by the fact that quantum states exhibiting probability transfer close to $\Delta_{\text{line}}$ are characterized by infinite spatial extent and infinite energy~\cite{YHH12}. Due to such factors, a laboratory demonstration of QB remains an open challenge\footnote{It has been proposed that QB could be observed in Bose-Einstein condensates~\cite{PTM13,MPM14}. Experimental realizations of an optical analog of QB have also been reported~\cite{EZB20,DGG22}.}.

QB becomes more pronounced and better amenable to experimental observation when considered in a ring, rather than on a line~\cite{Gou21}. More concretely, one considers a particle of mass $\mu$ and electric charge $q$ moving in a ring of radius $R$. The ring is pierced (normally to the plane of the ring) by a constant uniform magnetic field $B$. The (normalized) wave function $\psi$ is here assumed to have a non-negative angular momentum: in this context, QB then manifests itself as the possibility for the probability current
\begin{equation}
j(\theta,t) = \frac{\hbar}{\mu R^2} \left[ \Imag \left( \psi^* \frac{\partial \psi}{\partial \theta} \right) - \frac{q R^2 B}{2 \hbar c} |\psi|^2 \right]
\label{j_ring}
\end{equation}
to be negative, at some $\theta$ and $t$, despite the particle's angular momentum being, with certainty, non-negative. The probability transfer $\Delta$ through a fixed point on the ring, say $\theta=0$, over a time interval $-T/2 < t < T/2$ is still given by Eq.~\eqref{Delta} but now with the current~\eqref{j_ring}. A numerical analysis has shown that~\cite{Gou21}
\begin{equation}
	\inf_{\psi} \Delta = \Delta_{\text{ring}} \simeq -0.116816 \,,
\label{Delta_ring}
\end{equation}
and that the backflow-optimizing state is achieved for $\hbar T / 4 \mu R^2 \simeq 1.163635$ and $B = 0$. Thus, QB in a ring can be over three times more pronounced than QB on a line. In addition to this, the backflow-optimizing state in a ring appears to have a finite energy (and, in view of the system's geometry, a finite spatial extent) \cite{Gou21}.

Therefore, an important question is whether there are physical systems characterized by a probability transfer $\Delta$ smaller than $\Delta_{\text{ring}}$. Most of the attention to date has been focused on one-dimensional systems, where the probability transfer $\Delta$ has always been found to be bounded from below~\cite{BM94,MB98,MB98_Dirac,EFV05,PGK06,YHH12,BCL17,Gou19,DT19,ALS19,Gou21}. In this work, we rather consider a two-dimensional system.

Two-dimensional QB has been studied for a charged particle moving in a uniform vertical magnetic field in the infinite $(x, y)$ plane, both in the commutative~\cite{Str12} and noncommutative~\cite{PPR20} cases. As was noted by Strange in~\cite{Str12}, a noteworthy feature of such a system as compared to one-dimensional systems is that it allows for degenerate energy eigenstates, namely Landau levels in~\cite{Str12}. In turn, superpositions of degenerate Landau levels yield time-independent local currents $j$. Furthermore, by carefully tuning the coefficients of such superpositions, negative azimuthal currents can be seen to arise from superpositions of positive-angular-momentum Landau levels, which is thus a manifestation of QB for such a system. Therefore, Landau levels allow for time-independent backflow currents. This feature is in stark contrast with the transient backflow currents obtained in one-dimensional systems, and is potentially promising regarding the yet-to-be-performed experimental observation of the elusive effect of QB.

The Landau levels that arise for a charged particle in the infinite $(x,y)$ plane~\cite{Str12} are infinitely degenerate. However, in an actual experiment the particle would be confined within a finite region of space, hence typically alleviating the degeneracy of the energy spectrum. Therefore, this naturally prompts the question: can superpositions of degenerate energy eigenstates still allow for time-independent backflow currents in other, possibly finite, two-dimensional systems?

To answer the latter question, in the present work we consider a two-dimensional system where a charged particle is confined to move on a disk of finite radius $R$ that is punctured at the center and pierced through the center and normally to the disk by a magnetic flux line. This particular system allows to analytically compute the eigenenergies and corresponding eigenstates. We demonstrate the existence of superpositions of degenerate eigenstates that yield time-independent local backflow currents that, in addition, appear to be unbounded. Furthermore, since actual measurements typically correspond to some kind of space- and time-averages, we show the existence of time-independent spatially integrated backflow currents as well. Finally, we also show that our system allows for an unbounded dimensionless probability transfer $\Delta$, which is in stark contrast with the bounded probability transfers [such as~\eqref{Delta_line} or~\eqref{Delta_ring}] obtained for one-dimensional systems. Our work hence extends the range of available physical platforms that can offer a practical advantage regarding the experimental observation of QB.

The paper is organized as follows. We first set up the problem in section~\ref{disk_sec}, before we formulate QB in section~\ref{QB_sec}. We then discuss in section~\ref{time_indep_sec} how both the local current and the probability transfer are unbounded from below: this can be done by considering superpositions of two degenerate energy eigenstates. Concluding remarks are finally drawn in section~\ref{ccl_sec}.


\section{Punctured disk pierced by a magnetic flux}
\label{disk_sec}
We consider a nonrelativistic structureless quantum particle of mass 
$\mu$ and electric charge $q$ confined on the punctured disk 
$\mathcal{D}$, of center the origin $O$ and radius $R>0$, defined by
\begin{equation}
\mathcal{D} \equiv \left\{ \bol{r} \in 
\mathbb{R}^3 \quad | \quad 0 < r < R \; \; \text{and} \; \; 0 \leqslant 
\theta < 2 \pi \right\} \, .
\label{punct_disk_defm}
\end{equation}
The particle is subjected to the vector potential
\begin{equation}
\bol{A}(\bol{r}) = \frac{\eta}{r} \, \bol{e}_{\theta} \, ,
\label{A_exprm}
\end{equation}
where $\eta$ is a fixed parameter whose dimension is (length $\times$ energy)/(electric charge). The magnetic field corresponding to this vector potential, $ \bol{B}(\bol{r}) = \bol{\nabla} \times \bol{A}(\bol{r})$, vanishes for any $\bol{r} \neq \bol{0}$. Such a situation can be realized by an ideal infinite solenoid, oriented along the $z$ axis, and that has a 
vanishing radius (see e.g.~\cite{Grif}). The dynamical state $\ket{\Psi(t)}$ of the particle at any time $t$ is, in the position representation, described by a wave function $\Psi(r,\theta,t)$ that obeys the time-dependent Schr\"odinger equation
\begin{equation}
i \hbar \frac{\partial}{\partial t} \Psi(r,\theta,t) = H \Psi(r,\theta,t)
\label{TDSE_def}
\end{equation}
where the Hamiltonian $H$ is given by
\begin{equation}
\begin{aligned}
H &= \frac{1}{2 \mu} \left[ \bol{p} - \frac{q}{c} 
\bol{A}(\bol{r}) \right]^2\\
 &= - \frac{\hbar^2}{2 \mu} \left[ \frac{\partial^2}{\partial r^2} + 
\frac{1}{r} \frac{\partial}{\partial r} + \frac{1}{r^2} \left( 
\frac{\partial}{\partial \theta} - i \beta \right)^2 \right] \, ,
\label{H_exprm}
\end{aligned}
\end{equation}
in terms of the dimensionless parameter $\beta$ defined by
\begin{equation}
\beta \equiv \frac{q \eta}{\hbar c} \, ,
\label{beta_def}
\end{equation}
with $c$ denoting the speed of light (in vacuum).

Owing to the time independence and azimuthal symmetry of  the 
Hamiltonian~\eqref{H_exprm}, and imposing periodic boundary conditions $\psi(r,0) = \psi(r,2\pi)$ in the azimuthal direction as well as Dirichlet boundary conditions $\lim_{r \to 0} \psi(r,\theta) = \psi(R,\theta) = 0$ in the radial direction, the normalized eigenstates that satisfy the time-independent Schr\"odinger equation are given by
\begin{subequations}
\begin{equation}
\psi_{mn}(r,\theta) = \phi_{mn}(r) \, e^{i m \theta} \, ,
\label{psi_mn_compact_expr}
\end{equation}
with
\begin{equation}
\phi_{mn}(r) = \frac{1}{R \sqrt{\pi}} \frac{J_{\lvert M \rvert} \left(\gamma_{\lvert M \rvert n} \frac{r}{R} \right)}{\left\lvert J_{\lvert M \rvert + 1}(\gamma_{\lvert M \rvert n}) \right\rvert} \, ,
\label{phi_mn_def}
\end{equation}
\label{psi_mn_expr}%
\end{subequations}
where $m$ is an arbitrary integer, $n$ is a positive integer and $M$ is an arbitrary nonzero real number, defined by
\begin{equation}
M \equiv m - \beta.
\label{M_def}
\end{equation}
The function $J_{\lvert M \rvert}$ in~\eqref{phi_mn_def} is the Bessel function of the first kind of positive order $\lvert M 
\rvert$ (hence irrespectively of whether $M$ itself is positive or negative), 
while $\gamma_{\lvert M \rvert n}$ denotes the $n$th zero of 
$J_{\lvert M \rvert}$, that is
\begin{equation}
J_{\lvert M \rvert} \left( \gamma_{\lvert M \rvert n} \right) = 0
\label{zeros_J_def}
\end{equation}
for any $n \geqslant 1$. To the eigenstates~\eqref{psi_mn_expr} correspond the eigenenergies $E_{mn}$ given by
\begin{equation}
E_{mn} = \frac{\hbar^2}{2 \mu R^2} \gamma_{\lvert M \rvert n}^2 \, .
\label{E_mn_def}
\end{equation}

The complete set of eigenstates~\eqref{psi_mn_expr} is orthonormal,
\begin{equation}
\int_{\mathcal{D}} dS \, \psi_{mn}^*(r,\theta) 
\, \psi_{m'n'}(r,\theta) = \delta_{mm'} \, \delta_{nn'} \, ,
\label{orthog_cond_psi_mn}
\end{equation}
where the asterisk $^*$ denotes complex conjugation and $\delta$ is the 
Kronecker delta. The dynamical state $\Psi(r,\theta,t)$ of the particle at any time $t$ can thus in full generality be expanded in terms of the eigenstates $\psi_{mn}$ as
\begin{equation}
\Psi(r,\theta,t) = \sum_{m\neq \beta} \, \sum_{n \geqslant 1} c_{mn} \, 
\psi_{mn}(r,\theta) \, e^{- i E_{mn} t / \hbar} \, ,
\label{expansion_Psi_psi_mnm}
\end{equation}
where the complex coefficients $c_{mn}$ must satisfy the normalization 
condition
\begin{equation}
\sum_{m\neq \beta} \, \sum_{n \geqslant 1} \left\lvert c_{mn} 
\right\rvert^2 = 1
\label{normalization_cond_c_mnm}
\end{equation}
but are otherwise arbitrary.


\section{Formulation of the quantum backflow problem}
\label{QB_sec}
The kinetic angular momentum $\bol{L}$ is defined by
\begin{equation}
\bol{L} \equiv \bol{r} \times \left[ \bol{p} - \frac{q}{c} 
\bol{A}(\bol{r}) \right] \, ,
\label{Lambda_def}
\end{equation}
and is purely vertical in the present case,
\begin{equation}
\bol{L} = L_z \bol{e}_z = - \hbar \left( i \, 
\frac{\partial}{\partial \theta} + \beta \right) \bol{e}_z \, .
\label{Lambda_expr}
\end{equation}
It admits the energy eigenstates~\eqref{psi_mn_expr} as eigenstates, with $L_z \psi_{mn} = M\hbar \, \psi_{mn}$. The sign of $M$ hence allows to assign a precise direction of motion (counterclockwise or clockwise) to the eigenstates $\psi_{mn}$.

The latter fact can also be viewed from the current. The probability current $\bol{j}$ corresponding to $\Psi$ is given by 
\begin{equation}
\bol{j} (\bol{r},t) \equiv \frac{1}{\mu} \Real \left\{ \Psi^* 
(\bol{r},t) \left[ \bol{p} - \frac{q}{c} \bol{A} (\bol{r}) \right] \Psi 
(\bol{r},t) \right\},
\label{proba_current_defm}
\end{equation}
which in polar coordinates, and in view of
Eq.~\eqref{A_exprm}, reads
\begin{equation}
\bol{j} (r,\theta,t) = j_{\text{r}} (r,\theta,t) \, \bol{e}_{r} + 
j_{\text{a}} (r,\theta,t) \, \bol{e}_{\theta} \, ,
\label{j_psi_gen_exprm}
\end{equation}
with
\begin{subequations}
\begin{align}
j_{\text{r}} (r,\theta,t) &= - \frac{\hbar}{\mu} \Real \left( i \Psi^* 
\frac{\partial \Psi}{\partial r} \right) \, , \label{polar_coord_current_defm_r} \\[0.2cm]
j_{\text{a}} (r,\theta,t) &= - \frac{\hbar}{\mu r} \left[ \Real 
\left( i \Psi^* \frac{\partial \Psi}{\partial \theta} \right) + \beta 
\left\lvert \Psi \right\rvert^2 \right] \label{polar_coord_current_defm_a}
\end{align}
\label{coord_j_psi_def}%
\end{subequations}
being the radial and azimuthal components, respectively, of the current. It is then easy to check that evaluating the current for $\Psi = \psi_{mn}$ yields
\begin{subequations}
\begin{align}
j_{\text{r}} (r,\theta,t) \Big|_{\Psi = \psi_{mn}} &= 0 \, , \label{j_r_eigen}\\[0.2cm]
j_{\text{a}} (r,\theta,t) \Big|_{\Psi = \psi_{mn}} &= \frac{M \hbar}{\mu r} \left\lvert \phi_{mn}(r) \right\rvert^2 \, . \label{j_a_eigen}
\end{align}
\label{coord_j_psi_mnm}%
\end{subequations}
This shows that the probability current $\bol{j}$ for the 
eigenstates~\eqref{psi_mn_expr} is purely azimuthal and
independent of the angle $\theta$.
Furthermore, this 
ensures that $\bol{j} \big|_{\Psi = \psi_{mn}}$ is oriented along $+ 
\bol{e}_{\theta}$ when $M>0$, and along $- \bol{e}_{\theta}$ when $M<0$. 

Similar to the case of a charged particle in a one-dimensional ring~\cite{Gou21} or in the infinite $(x,y)$ plane~\cite{Str12,PPR20}, our formulation of QB is based on considering quantum states $\Psi$ that contain only eigenstates $\psi_{mn}$ with positive kinetic angular momentum, \textit{i.e.} we only include $m > \beta$ in the expansion~\eqref{expansion_Psi_psi_mnm}. Therefore, in the sequel we restrict to the class of states given by
\begin{equation}
\Psi(r,\theta,t) = \sum_{m > \beta} \, \sum_{n \geqslant 1} c_{mn} \, 
\psi_{mn}(r,\theta) \, e^{- i E_{mn} t / \hbar} \, ,
\label{expansion_Psi_psi_mn_pos_Mm}
\end{equation}
with the normalization condition~\eqref{normalization_cond_c_mnm} now 
reading
\begin{equation}
\sum_{m > \beta} \, \sum_{n \geqslant 1} \left\lvert c_{mn} 
\right\rvert^2 = 1 \, .
\label{normalization_cond_c_mn_pos_Mm}
\end{equation}
In view of~\eqref{coord_j_psi_mnm}, the states of the 
form~\eqref{expansion_Psi_psi_mn_pos_Mm} hence correspond to superpositions of 
states $\psi_{mn}$ that all individually exhibit a positive azimuthal 
current, $j_{\text{a}} \big|_{\Psi = \psi_{mn}} > 0$.

QB then occurs whenever the azimuthal component $j_{\text{a}}$ of the probability current associated with the state \eqref{expansion_Psi_psi_mn_pos_Mm} takes on negative values. In fact, there are three (closely related) quantities that can be used to quantify QB. Due to the rotational symmetry of the system, it is sufficient to introduce these quantities only for these points of the disk that lie on the ray $\theta = 0$.

First, the azimuthal probability current $j_{\text{a}}$ quantifies the strength of QB locally in space and time. Thus, QB occurs at a spatial point $(r,0)$ and a time $t$ whenever $j_{\text{a}}(r,0,t) < 0$, and the smaller $j_{\text{a}}$ the more pronounced QB at this particular space-time point.

Second, the spatially integrated current
\begin{equation}
\mathcal{J}(r_1, r_2, t) = \int_{r_1}^{r_2} dr \, j_{\text{a}}(r, 0, t)
\label{sp_int_curr_defn}
\end{equation}
is the rate of probability transfer through the straight radial section of the disk connecting the points $(r_1,0)$ and $(r_2,0)$, with $0 < r_1 < r_2 \leqslant R$, at a fixed time $t$. Clearly, the condition $\mathcal{J}(r_1, r_2, t) < 0$ is stronger than the condition $j_{\text{a}}(r,0,t) < 0$ for some $r_1 < r < r_2$; indeed, the former implies the latter, but not the other way around.

Third,
\begin{equation}
\Delta(r_1, r_2, T) = \int_{-T/2}^{T/2} dt \, \mathcal{J}(r_1, r_2, t)
\label{Delta_def}
\end{equation}
is the total probability transfer through the straight section between $(r_1,0)$ and $(r_2,0)$ over the time interval $-T/2 < t < T/2$. The condition $\Delta(r_1, r_2, T) < 0$ is stronger than $\mathcal{J}(r_1, r_2, t) < 0$ for some $-T/2 < t < T/2$.

The dimensionless quantity $\Delta$ defined by~\eqref{Delta_def} is a two-dimensional analog of the probability transfer typically addressed in QB studies in one-dimensional settings [cf.~Eq.~\eqref{Delta}]. Remarkably, it behaves dramatically differently than its one-dimensional couterparts: while, in one dimension, the latter appear to be bounded from below~\cite{BM94,MB98,MB98_Dirac,EFV05,PGK06,YHH12,BCL17,Gou19,DT19,ALS19,Gou21}, the two-dimensional probability transfer~\eqref{Delta_def} is \textit{unbounded from below}. This can be seen by considering a state that exhibits a time-independent backflow current, as we now discuss.


\section{Time-independent backflow current}
\label{time_indep_sec}

An important facet of QB in one-dimensional systems is the fact that the probability current $j$, given e.g. by Eq.~\eqref{j_line} in the line case or Eq.~\eqref{j_ring} in the ring case, can stay appreciably negative only during a finite time. The situation appears to be drastically different in two dimensions: indeed, as was noted by Strange for an electron in a constant magnetic field~\cite{Str12}, the local current $j_{\text{a}}$ can remain negative indefinitely.

To show that the latter fact remains true in our (finite) system it is sufficient to consider states $\Psi$ comprised of only two eigenstates $\psi_{mn}$ and $\psi_{m'n'}$ with, in accordance with \eqref{expansion_Psi_psi_mn_pos_Mm}, $m > \beta$, $m' > \beta$, $n \geqslant 1$, $n' \geqslant 1$, and $(m,n) \neq (m', n')$. Furthermore, hereinafter we assume that $\psi_{mn}$ and $\psi_{m'n'}$ have the same energy, i.e.
\begin{equation}
E_{mn} = E_{m'n'} \, .
\label{degeneracy_condition_E}
\end{equation}
In view of Eq.~\eqref{E_mn_def} [and remembering the definition~\eqref{M_def}], this degeneracy condition is equivalent to the requirement that the $n\text{th}$ zero of the Bessel function $J_{m-\beta}$ coincides with the $n'\text{th}$ zero of $J_{m'-\beta}$ (which is not forbidden as long as $m - \beta$ is irrational, see e.g.~\cite{Watson}), i.e.
\begin{equation}
	\gamma_{m-\beta,n} = \gamma_{m'-\beta,n'} \,.
\label{degeneracy_condition}
\end{equation}
In view of~\eqref{expansion_Psi_psi_mn_pos_Mm} and~\eqref{degeneracy_condition_E}, the state $\Psi$ that we consider in the sequel hence reads
\begin{equation}
\Psi(r,\theta,t) = \left[ c_{mn} \, \psi_{mn}(r,\theta) + c_{m'n'} \, \psi_{m'n'}(r,\theta) \right] e^{- i E_{mn} t / \hbar} \, ,
\label{Psi_for_2_states}
\end{equation}
subject to the normalization condition
\begin{equation}
	|c_{mn}|^2 + |c_{m'n'}|^2 = 1 \, .
\label{norm_for_2_states}
\end{equation}

The degeneracy condition~\eqref{degeneracy_condition_E}-\eqref{degeneracy_condition} is the crucial difference that distinguishes the two-dimensional system studied here from all the previous (one-dimensional) systems that have been considered to this date regarding the maximal QB, for which degeneracy is \textit{not} possible. We also note that the degeneracy condition~\eqref{degeneracy_condition} requires the presence of an external magnetic flux: indeed, if $\beta = 0$, the orders $m - \beta$ and $m' - \beta$ of the Bessel functions become integers, and distinct Bessel functions of integral orders are known to have no zeros in common (which is the so-called Bourget hypothesis, see e.g.~\cite{Watson}).


\subsection{Spatially local current \texorpdfstring{$j_{\text{a}}$}{j}}
\label{subsec:local_current}
Substituting Eqs.~\eqref{Psi_for_2_states} and \eqref{psi_mn_compact_expr} into Eq.~\eqref{polar_coord_current_defm_a} immediately shows that the azimuthal local current $j_{\text{a}}$ is time independent and we have
\begin{align}
j_{\text{a}}(r,0) &= \frac{\hbar}{\mu r} \bigg[ |c_{mn}|^2 M \phi_{mn}^2(r) + |c_{m'n'}|^2 M' \phi_{m'n'}^2(r) \nonumber \\ &+ \Real \left\{ c_{mn}^* c_{m'n'} \right\} (M+M') \phi_{mn}(r) \phi_{m'n'}(r)  \bigg] \, ,
\label{j_a_t-indep}
\end{align}
where, by definition, $M = m - \beta$ [cf.~Eq.~\eqref{M_def}] and, similarly, $M' = m' - \beta$.

Our aim is now i) to find the smallest possible value of the azimuthal current~\eqref{j_a_t-indep}, and ii) to demonstrate that this smallest value can be arbitrarily negative. The minimum of $j_{\text{a}}$ over the space of all normalized expansion coefficients $c_{mn}$ and $c_{m'n'}$ is given by (see appendix~\ref{app:minimization_of_currents} for the derivation)
\begin{widetext}
\begin{subequations}
\begin{align}
	\min j_{\text{a}} & = \frac{\hbar}{2 \mu r} \left\{ M \phi_{mn}^2(r) + M' \phi_{m'n'}^2(r) - \sqrt{\big[ M \phi_{mn}^2(r) + M' \phi_{m'n'}^2(r) \big]^2 + (M - M')^2 \phi_{mn}^2(r) \phi_{m'n'}^2(r)} \right\} \label{min_j_a_b} \\
	&= \frac{\hbar}{2 \mu r} \left\{ M \phi_{mn}^2(r) + M' \phi_{m'n'}^2(r) - \sqrt{\big[ M^2 \phi_{mn}^2(r) + M'^2 \phi_{m'n'}^2(r) \big] \big[ \phi_{mn}^2(r) + \phi_{m'n'}^2(r) \big]} \right\} \label{min_j_a_c} \,.
\end{align}
\label{min_j_a}
\end{subequations}
\end{widetext}

First, it is clear from~\eqref{min_j_a_b} that $\min j_{\text{a}} \leqslant 0$ for any $r$: the azimuthal current~\eqref{j_a_t-indep} can thus indeed be negative for states of the form~\eqref{Psi_for_2_states}. Furthermore, we now argue that $\min j_{\text{a}}$ is unbounded from below (technical details can be found in appendix~\ref{unb_min_j_a_app}). To this end, let's consider some particular value(s) of $r$ for which
\begin{equation}
\phi_{m'n'}^2(r) \sim \phi_{mn}^2(r) \sim \mathcal{O} (1) \, .
\label{phi_prime_order_phi_order_1}
\end{equation}
Then, if we write $M' = u M$, we get upon combining~\eqref{min_j_a_c} with~\eqref{phi_prime_order_phi_order_1}
\begin{equation}
\min j_{\text{a}} \sim \frac{\hbar M \phi_{mn}^2(r)}{2 \mu r} \left[ 1 + u - \sqrt{2(1 + u^2)} \right] \, .
\label{min_j_a_approx}
\end{equation}
In the limit $u \to \infty$ we then clearly see that
\begin{equation}
\min j_{\text{a}} \sim \frac{\hbar M \phi_{mn}^2(r)}{2 \mu r} \left[ 1 + \left( 1 - \sqrt{2} \right) u \right] \underset{u \to \infty}{\to} - \infty \, .
\label{min_j_a_lim}
\end{equation}

This hence demonstrates that, by properly choosing the two degenerate eigenstates $\phi_{mn}(r)$ and $\phi_{m'n'}(r)$, one can engineer an arbitrarily small current locally. (We note that the local backflow current has also been found to be unbounded in the one-dimensional case of a free particle on a line~\cite{BM94,BCL17}).


\subsection{Spatially integrated current \texorpdfstring{$\mathcal{J}$}{J}}
\label{spat_int_cur_subsec}

We now construct the spatially integrated current $\mathcal{J}$, as defined by~\eqref{sp_int_curr_defn}, that is associated with the (time-independent) azimuthal current~\eqref{j_a_t-indep}, and we get
\begin{align}
\mathcal{J}(r_1,r_2) &= \frac{\hbar}{\mu R^2} \bigg[ |c_{mn}|^2 M S_{mnmn} 
+ |c_{m'n'}|^2 M' S_{m'n'm'n'} \nonumber \\ &+ \Real \left\{ 
c_{mn}^* c_{m'n'} \right\} (M+M') S_{mnm'n'}  \bigg] 
\, ,
\label{int_j_a_t-indep}
\end{align}
where
\begin{equation}
S_{mnm'n'} = R^2 \int_{r_1}^{r_2} \frac{dr}{r}\phi_{mn}(r)\phi_{m'n'}(r) \, .
\label{S_mn_def}
\end{equation}
Following the method outlined in appendix~\ref{app:minimization_of_currents}, we can here again find the smallest possible value $\min \mathcal{J}$ of~\eqref{int_j_a_t-indep} over the space of all normalized expansion coefficients $c_{mn}$ and $c_{m'n'}$, and we get
\begin{widetext}
\begin{subequations}
\begin{align}
\min \mathcal{J} &= \frac{\hbar}{2 \mu R^2} \left( M S_{mnmn}+ M' S_{m'n'm'n'} - \sqrt{\big[ M S_{mnmn}- M' S_{m'n'm'n'} \big]^2 + (M + M')^2 S_{mnm'n'}^2} \right) \label{min_int_j_a_a} \\
&= \frac{\hbar}{2 \mu R^2} \left( M S_{mnmn}+ M' S_{m'n'm'n'} - \sqrt{\big[ M S_{mnmn}+ M' S_{m'n'm'n'} \big]^2 + \rho} \right) \label{min_int_j_a_b} \, ,
\end{align}
\label{min_int_j_a}
\end{subequations}
\end{widetext}
where $\rho$ is defined by
\begin{equation}
\rho = (M + M')^2 S_{mnm'n'}^2 - 4M M' S_{mnmn}S_{m'n'm'n'} \, .
\label{rho_def}
\end{equation}

Note that the quantity $\rho$ given by~\eqref{rho_def} can take on both positive and negative values depending on the values of the parameters $(m, n, m', n', \beta, r_1, r_2)$ (as we explicitly observe numerically). Therefore, as is clear from~\eqref{min_int_j_a_b}, $\min \mathcal{J}$ is only negative when $\rho$ is positive: this is for instance the case for $m=1, n=3, m'=6, n'=1, \beta \approx 0.69117, r_1 = 3R/10, r_2 = 7R/10$, for which we indeed get $\min \mathcal{J} < 0$. However, we have not been able to find an argument that would allow us to conclude on either the boundedness or unboundedness of $\min \mathcal{J}$.


\subsection{Probability transfer \texorpdfstring{$\Delta$}{Delta}}

Finally, we now show that the probability transfer $\Delta$ is unbounded from below. Since the integrated current~\eqref{int_j_a_t-indep} is by construction time independent, we hence readily get from the definition~\eqref{Delta_def} of $\Delta$ that
\begin{equation}
\Delta(r_1, r_2, T) = T \mathcal{J}(r_1, r_2) \, .
\label{Delta_expr}
\end{equation}
We now choose the parameters $(m, n, m', n', \beta, r_1, r_2)$ such that $\mathcal{J}(r_1, r_2) < 0$ (which, as mentioned in section~\ref{spat_int_cur_subsec} above, can occur for $m~=~1, n=3, m'=6, n'=1, \beta \approx 0.69117, r_1 = 3R/10, r_2 = 7R/10$). It is then immediate that
\begin{equation}
\lim\limits_{T \to \infty} \Delta(r_1, r_2, T) = - \infty \, .
\label{Delta_unbounded}
\end{equation}

Therefore, it is clear from~\eqref{Delta_expr}-\eqref{Delta_unbounded} that $\Delta$ can be arbitrarily negative: as soon as $\mathcal{J}(r_1, r_2) < 0$, one then simply has to wait for a long enough time $T$. This is in particular a drastic difference with the (one-dimensional) ring case~\cite{Gou21}, where QB becomes weaker at large $T$.


\section{Conclusion and outlook}
\label{ccl_sec}

We formulated and studied QB for a charged particle on a punctured disk. We saw that for such a finite two-dimensional system two quantifiers of QB, namely the local azimuthal current~\eqref{polar_coord_current_defm_a} as well as the dimensionless total probability transfer~\eqref{Delta_def}, are unbounded from below. This is in particular in stark contrast with the bounded probability transfers obtained to this date for one-dimensional systems. Similarly to what Strange noticed for Landau levels in~\cite{Str12}, this remarkable feature is seen to arise from the existence of superpositions of degenerate energy eigenstates [Eq.~\eqref{Psi_for_2_states}] that give rise to time-independent backflow currents. Such time-independent backflow currents never arise in one-dimensional systems, where backflow currents are transient. Therefore, we believe that this makes two-dimensional systems in general, and the punctured disk studied in this work in particular, promising physical platforms that present a distinct practical advantage in view of the yet-to-be-performed experimental observation of the elusive phenomenon of QB.

While we have seen that degeneracy allows for an unbounded probability transfer $\Delta$ as $T \to \infty$ [Eq.~\eqref{Delta_unbounded}], it would now be interesting to study whether $\Delta$ remains unbounded even for finite times $T$. An other possible follow-up question could be to investigate whether the unboundedness of $\Delta$ actually requires degeneracy: in other words, is $\Delta$ still unbounded for superpositions of eigenstates of different energies, and thus for time-dependent currents?


\section*{Acknowledgements}

M. B. is grateful to M. T. Eiles for valuable discussions.


\appendix


\section{Derivation of \texorpdfstring{Eqs.~\eqref{min_j_a_b} and~\eqref{min_int_j_a_a}}{}}
\label{app:minimization_of_currents}

Both the spatially local and spatially integrated currents, Eqs.~\eqref{j_a_t-indep} and \eqref{int_j_a_t-indep} respectively, have the same dependence on the expansion coefficients $c_{mn}$ and $c_{m'n'}$:
\begin{equation}
	f = |c_{mn}|^2 A + |c_{m'n'}|^2 B + \Real \left\{ c_{mn}^* c_{m'n'} \right\} C \,,
	\label{f_def_app}
\end{equation}
where $A$, $B$, and $C$ are constants. In order to find the minimum of $f$ over the space of all normalized expansion coefficients, we parametrize the latter as 
\begin{equation}
	c_{mn} = \cos \frac{\varphi}{2} \,, \qquad c_{m'n'} = e^{i \gamma} \sin \frac{\varphi}{2} \,,
	\label{parametrization_def}
\end{equation}
with $\varphi\in [0, \pi]$ and $\gamma \in [0, 2 \pi)$. This yields
\begin{align*}
	f &= A \cos^2 \frac{\varphi}{2} + B \sin^2 \frac{\varphi}{2} + C \cos \frac{\varphi}{2} \sin \frac{\varphi}{2} \cos \gamma \\ &= \frac{A + B}{2} + \frac{A - B}{2} \cos \varphi + \frac{C}{2} \sin \varphi \cos \gamma \,.
\end{align*}
Minimizing $f$ with respect to $\gamma$, we find
\begin{equation*}
	\min_{\gamma} f = \frac{A + B}{2} + \frac{A - B}{2} \cos \varphi - \frac{|C|}{2} \sin \varphi \,.
\end{equation*}
In order to minimize the last expression with respect to $\varphi$, we perform the following transformation:
\begin{align*}
	&(A-B) \cos \varphi - |C| \sin \varphi \\ &= \sqrt{(A-B)^2 + C^2} ( \cos \varphi \cos \varphi_0 - \sin \varphi \sin \varphi_0 ) \\
	&= \sqrt{(A-B)^2 + C^2} \cos(\varphi + \varphi_0) \,,
\end{align*}
where $\varphi_0 \in [0, \pi]$ is defined by
\begin{equation*}
	\cos \varphi_0 = \frac{A-B}{\sqrt{(A-B)^2 + C^2}} \, , \, \sin \varphi_0 = \frac{|C|}{\sqrt{(A-B)^2 + C^2}} \,.
\end{equation*}
Then, the function
\begin{equation*}
	\min_{\gamma} f = \frac{1}{2} \left[ A + B + \sqrt{(A-B)^2 + C^2} \cos(\varphi + \varphi_0) \right]
\end{equation*}
has a minimum at $\varphi = \pi - \varphi_0$, and
\begin{equation}
	\min_{\varphi, \gamma} f = \frac{A + B - \sqrt{(A-B)^2 + C^2}}{2} \,.
	\label{min_f_expr}
\end{equation}
Substituting the adequate values of $A,B$ and $C$ into~\eqref{min_f_expr} and performing a few straightforward algebraic manipulations then readily yields the expressions~\eqref{min_j_a_b} and~\eqref{min_int_j_a_a} of $\min j_{\text{a}}$ and $\min \mathcal{J}$, respectively.


\section{Unboundedness of \texorpdfstring{$\min j_{\text{a}}$}{min j}}
\label{unb_min_j_a_app}

In this appendix we demonstrate the unboundedness of the spatially local current $j_{\text{a}}$, as described by Eqs.~\eqref{phi_prime_order_phi_order_1}-\eqref{min_j_a_lim}. We recall that we consider here couples of integers $(m,n)$ and $(m',n')$ such that $m > \beta$, $m' > \beta$, $n \geqslant 1$, $n' \geqslant 1$, and $(m,n) \neq (m', n')$: therefore, we have here
\begin{equation}
M \equiv m - \beta > 0 \qquad \text{and} \qquad M' \equiv m' - \beta > 0 \, .
\label{M_Mpr_def}
\end{equation}
Furthermore, we assume the degeneracy condition~\eqref{degeneracy_condition_E}, i.e.
\begin{equation}
E_{mn} = E_{m'n'} \, ,
\label{degeneracy_condition_E_supp_mat}
\end{equation}
that is for the zeros $\gamma_{m-\beta,n}$ and $\gamma_{m'-\beta,n'}$ [Eq.~\eqref{degeneracy_condition}]
\begin{equation}
\gamma_{m-\beta,n} = \gamma_{m'-\beta,n'} \, .
\label{degeneracy_condition_supp_mat}
\end{equation}
For completeness, we also recall that we have [Eq.~\eqref{phi_mn_def}]
\begin{equation}
\phi_{mn}(r) = \frac{1}{R \sqrt{\pi}} \frac{J_{M} \left(\gamma_{M n} \frac{r}{R} \right)}{\left\lvert J_{M + 1}(\gamma_{M n}) \right\rvert}
\label{phi_mn_alt_expr}
\end{equation}
and
\begin{equation}
\phi_{m'n'}(r) = \frac{1}{R \sqrt{\pi}} \frac{J_{M'} \left(\gamma_{M' n'} \frac{r}{R} \right)}{\left\lvert J_{M' + 1}(\gamma_{M' n'}) \right\rvert} \, ,
\label{phi_mPrnPr_alt_expr}
\end{equation}
as well as the minimal current [Eq.~\eqref{min_j_a_c}]
\begin{align}
\min j_{\text{a}} = \frac{\hbar}{2 \mu r} \biggl\{ M \phi_{mn}^2(r) + M' \phi_{m'n'}^2(r) \nonumber \\[0.2cm]
- \sqrt{\big[ M^2 \phi_{mn}^2(r) + M'^2 \phi_{m'n'}^2(r) \big] \big[ \phi_{mn}^2(r) + \phi_{m'n'}^2(r) \big]} \biggr\} \, .
\label{min_j_a_c_supp_mat}
\end{align}

\vspace{0.5cm}

We first introduce the parameter $u$ defined through
\begin{equation}
M' \equiv u M \, ,
\label{u_def}
\end{equation}
as well as the function $v(r)$ defined by
\begin{equation}
\phi_{m'n'}^2 (r) \equiv v(r) \phi_{mn}^2 (r) \, .
\label{v_def}
\end{equation}
We then rewrite~\eqref{min_j_a_c_supp_mat} in terms of these parameters $u$ and $v$ , and we get
\begin{align}
\min j_{\text{a}} = \frac{\hbar M}{2 \mu r} \phi_{mn}^2(r) \biggl\{ 1 + u v(r) \nonumber \\[0.2cm]
- \sqrt{\left[ 1 + u^2 v(r) \right] \left[ 1 + v(r) \right]} \biggr\} \, .
\label{min_j_a_u_v_expr_app}
\end{align}

We now derive a relevant upper bound for $\min j_{\text{a}}$. Since we know from the definitions~\eqref{M_Mpr_def} and~\eqref{u_def}-\eqref{v_def} that [the case $v(r)=0$ is ruled out since in this case $\min j_{\text{a}}=0$]
\begin{equation}
u > 0 \quad \text{and} \quad v > 0 \, ,
\label{u_v_pos}
\end{equation}
we then clearly have
\begin{align*}
\sqrt{\left[ 1 + u^2 v(r) \right] \left[ 1 + v(r) \right]} = \sqrt{1 + v(r) + u^2 v(r) + u^2 v^2(r)} \\[0.3cm]
> \sqrt{u^2 v(r) + u^2 v^2(r)} = u v(r) \sqrt{1 + \frac{1}{v(r)}} \, ,
\end{align*}
that is
\begin{equation*}
- \sqrt{\left[ 1 + u^2 v(r) \right] \left[ 1 + v(r) \right]} < - u v(r) \sqrt{1 + \frac{1}{v(r)}} \, ,
\end{equation*}
and thus
\begin{align}
1 + u v(r) - \sqrt{\left[ 1 + u^2 v(r) \right] \left[ 1 + v(r) \right]} \nonumber \\[0.3cm]
< 1 + u v(r) - u v(r) \sqrt{1 + \frac{1}{v(r)}} \, .
\label{sqrt_ineq}
\end{align}
Therefore, combining~\eqref{min_j_a_u_v_expr_app} with~\eqref{sqrt_ineq} yields the following inequality:
\begin{equation}
\min j_{\text{a}} < \frac{\hbar M}{2 \mu r} \phi_{mn}^2(r) \left\{ 1 + u v(r) \left[ 1 - \sqrt{1 + \frac{1}{v(r)}} \right] \right\} \, ,
\label{min_j_a_ineq}
\end{equation}
which we emphasize is valid \textit{for any $r$ where $v(r) \neq 0$}.

Our aim is now to demonstrate that the right-hand side in the inequality~\eqref{min_j_a_ineq} can achieve arbitrarily negative values at some well-chosen values of $r$. To this end, we first note that in view of~\eqref{u_v_pos} we have
\begin{equation*}
\sqrt{1 + \frac{1}{v(r)}} > 1 \, ,
\end{equation*}
so that
\begin{equation}
u v(r) \left[ 1 - \sqrt{1 + \frac{1}{v(r)}} \right] < 0 \, .
\label{uv_term_neg}
\end{equation}
Therefore, this a priori allows the right-hand side in~\eqref{min_j_a_ineq} to indeed be negative. A possible route could thus be to take the limit $u \to \infty$: this could make the right-hand side in~\eqref{min_j_a_ineq} decrease to $-\infty$, at least at some particular, well-chosen values of $r$. However, to demonstrate that this is indeed the case requires us to be careful about two things:

\begin{enumerate}[i)]
\item the value that $\phi_{mn}^2(r)$ [which is a global factor in the right-hand side of~\eqref{min_j_a_ineq}] takes on at these particular values of $r$ that we choose: indeed, $\phi_{mn}^2(r)$ must take on a \textit{finite} (i.e. non negligible) value for the right-hand side in~\eqref{min_j_a_ineq} to possibly go to $-\infty$;

\item the value of $v(r)[ 1 - \sqrt{1 + 1/v(r)} ]$ at these particular values of $r$: this term must also be \textit{finite} so as not to compensate the eventual growth of $u$.
\end{enumerate}

Our strategy is thus as follows. First, we identify the values of $r$ where $\phi_{mn}^2(r)$ reaches a local maximum, hence ensuring in particular that $\phi_{mn}^2(r)$ takes on finite values. This task can be tackled analytically by considering the regime $u \gg 1$, which allows us to rewrite $\phi_{mn}^2(r)$ by means of asymptotic expansions of Bessel functions: this is discussed in section~\ref{asympt_subsec}. We then argue in section~\ref{brack_qty_subsec} that $v(r)[ 1 - \sqrt{1 + 1/v(r)} ]$ remains finite at these particular values of $r$. We conclude in section~\ref{unb_min_j_a_subsec}.


\subsection{Asymptotic expansion of \texorpdfstring{$\phi_{mn}(r)$}{phi}}
\label{asympt_subsec}

From now on, we assume that the parameter $u$ is very large,
\begin{equation}
u \gg 1 \, .
\label{large_u}
\end{equation}
Table~\ref{table1} below contains numerical evidence that it is indeed a priori possible to find values of the parameters $(m,n,m',n',\beta)$ that i) satisfy the degeneracy condition~\eqref{degeneracy_condition_supp_mat} and ii) yield values of $u$ of increasing orders of magnitude, namely from $10$ to $10^4$ here. Here we restricted our attention to $m=n'=1$: this allowed us to perform a systematic search for all the corresponding parameters $n,m'$ and $\beta$ that yield common zeros [i.e. such that~\eqref{degeneracy_condition_supp_mat} is satisfied]. We observed in this case that to any given $n$ corresponds \textit{at most} a unique couple of values $(m', \beta)$ that makes the degeneracy condition~\eqref{degeneracy_condition_supp_mat} satisfied. We have searched for all integers $n$ between 3 and 164, and identified all the corresponding relevant values of $m'$ (that take values between 6 and 501). Since such a numerical search is fundamentally bound to identify finite values, we hence assume that it remains possible to find arbitrarily large values of $n$ and $m'$ that will still satisfy the degeneracy condition~\eqref{degeneracy_condition_supp_mat}: this then ensures the validity of the regime~\eqref{large_u}.


\begin{table*}[ht]
\centering
\caption{Values of $(m,n)$ and $(m',n')$ yielding increasing values of $u$. The indicated values of $\beta$ ensure the validity of both the degeneracy condition~\eqref{degeneracy_condition} and of $M' = u M$.}
\vskip 0.2 cm
\begin{tabular}{ |M{0.7cm}|M{1cm}|M{1cm}|M{0.7cm}|M{2.5cm}|M{1.5cm}|N }
\hline
\vspace{0.1cm}
$m$ & $n$ & $m'$ & $n'$ & $\beta$ & $u$ & \\[0.1cm]
\hline\hline
\vspace{0.1cm}
1 & 3 & 6 & 1 & 0.691169346793 & 17 & \\[0.25cm]
1 & 9 & 23 & 1 & 0.8494862493541 & 147 & \\[0.25cm]
1 & 28 & 80 & 1 & 0.957622423454 & 1865 & \\[0.25cm]
1 & 102 & 308 & 1 & 0.973770549551 & 11705 & \\[0.1cm]
\hline
\end{tabular}
\label{table1}
\end{table*}


In view of the definition~\eqref{u_def} of $u$, our assumption~\eqref{large_u} is hence valid for
\begin{equation}
M \; \text{fixed} \quad \text{and} \quad M' \gg 1 \, .
\label{M_Mprime_hyp}
\end{equation}
Let's now quickly discuss the impact of our assumption~\eqref{large_u} on the degeneracy condition~\eqref{degeneracy_condition_supp_mat}. In view of~\eqref{M_Mprime_hyp}, it is clear that the zero $\gamma_{M'n'}$ is thus a zero of the Bessel function $J_{M'}$ of a large order. Therefore, since it is well known (see~\cite{Watson}, pp. 485-486) that
\begin{equation}
\gamma_{M'n'} > M' \quad , \quad \forall n' \, ,
\label{zero_larger_order}
\end{equation}
the degeneracy condition~\eqref{degeneracy_condition_supp_mat} hence reads here
\begin{equation}
\gamma_{M'n'} = \gamma_{Mn} > M' \gg 1 \, .
\label{large_zero}
\end{equation}
The latter hence also requires, because $M$ is fixed according to~\eqref{M_Mprime_hyp}, that
\begin{equation}
n \gg 1 \, .
\label{n_large}
\end{equation}

The fact that $M$ is fixed and $\gamma_{Mn} \gg 1$ now suggests to rewrite $\phi_{mn}(r)$ [as given by~\eqref{phi_mn_alt_expr}] by means of asymptotic expansions for Bessel functions of finite orders and large arguments. While this can be done without further assumption for $J_{M + 1}(\gamma_{M n})$, to expand $J_{M} (\gamma_{M n} r/R )$ requires to assume that $r/R$ is not negligible. Therefore, in the sequel we only consider the portion of the disk away from the direct vicinity of the center of the disk, i.e. we consider
\begin{equation}
r \in (R_1,R) \, ,
\label{r_finite}
\end{equation}
where $R_1$ is finite (for concreteness, we may for instance have in mind $R_1 = R/2$). [The assumption~\eqref{r_finite} is harmful for what we want to do here. Indeed, our numerical investigations strongly suggest anyway that $\min j_{\text{a}}$ achieves its most negative values for finite values of $r$, and for values of $r$ that are closer and closer to $R$ as we increase $M'$.]

We now use the well-known asymptotic expansion of a Bessel function of a fixed, finite order and very large argument~\cite{GradRyz,AbrSteg}, and we have here
\begin{equation}
J_{M} \left(\gamma_{M n} \frac{r}{R} \right) \sim \sqrt{\frac{2 R}{\pi \gamma_{M n} r}} \, \cos \left( \gamma_{M n} \frac{r}{R} - \frac{M \pi}{2} - \frac{\pi}{4} \right)
\label{asympt_Bessel_1}
\end{equation}
and
\begin{equation}
J_{M+1} \left(\gamma_{M n} \right) \sim \sqrt{\frac{2}{\pi \gamma_{M n}}} \, \cos \left[ \gamma_{M n} - \frac{(M+1) \pi}{2} - \frac{\pi}{4} \right] \, .
\label{asympt_Bessel_2}
\end{equation}
Noting that
\begin{equation*}
\cos \left[ \gamma_{M n} - \frac{(M+1) \pi}{2} - \frac{\pi}{4} \right] = \sin \left( \gamma_{M n} - \frac{M \pi}{2} - \frac{\pi}{4} \right) \, ,
\end{equation*}
we hence get for~\eqref{asympt_Bessel_2}
\begin{equation}
J_{M+1} \left(\gamma_{M n} \right) \sim \sqrt{\frac{2}{\pi \gamma_{M n}}} \, \sin \left( \gamma_{M n} - \frac{M \pi}{2} - \frac{\pi}{4} \right) \, .
\label{asympt_Bessel_3}
\end{equation}
Substituting the asymptotic expansions~\eqref{asympt_Bessel_1} and~\eqref{asympt_Bessel_3} into~\eqref{phi_mn_alt_expr} hence yields the following asymptotic expansion of $\phi_{mn}(r)$:
\begin{equation}
\phi_{mn}(r) \sim \frac{1}{\sqrt{R \pi r}} \frac{\cos \left( \gamma_{M n} \frac{r}{R} - \frac{M \pi}{2} - \frac{\pi}{4} \right)}{\left\lvert \sin \left( \gamma_{M n} - \frac{M \pi}{2} - \frac{\pi}{4} \right) \right\rvert} \, .
\label{phi_mn_asympt}
\end{equation}

Now, since our aim is to identify some convenient values of $r$ where $\phi_{mn}(r)$ takes on finite values, we merely find the location of the extrema of $\phi_{mn}$. This is actually easily done by using the asymptotic expansion~\eqref{phi_mn_asympt}: indeed, since $\gamma_{M n} r/R \gg 1$ here, the extrema of $\phi_{mn}(r)$ can be approximated to be the extrema of the cosine function in the numerator in~\eqref{phi_mn_asympt} (this can be e.g. seen from the function $\cos (x)/\sqrt{x}$, and studying the equation that gives the stationary points of the latter function for large $x$). Therefore, let's denote by $r_k$ these particular values of $r$ where the numerator of $\phi_{mn}(r)$ takes on its extremal values. These are determined by the condition
\begin{equation*}
\gamma_{M n} \frac{r_k}{R} - \frac{M \pi}{2} - \frac{\pi}{4} = k \pi \quad , \quad k \in \mathbb{Z} \, ,
\end{equation*}
hence yielding
\begin{equation}
r_k = \frac{R}{\gamma_{M n}} \left( k + \frac{M}{2} + \frac{1}{4} \right) \pi \quad , \quad k \in \mathbb{Z} \, . 
\label{r_k_expr}
\end{equation}

\begin{center}
\begin{figure}[!ht]
\centering
\includegraphics[width=1.0\linewidth]{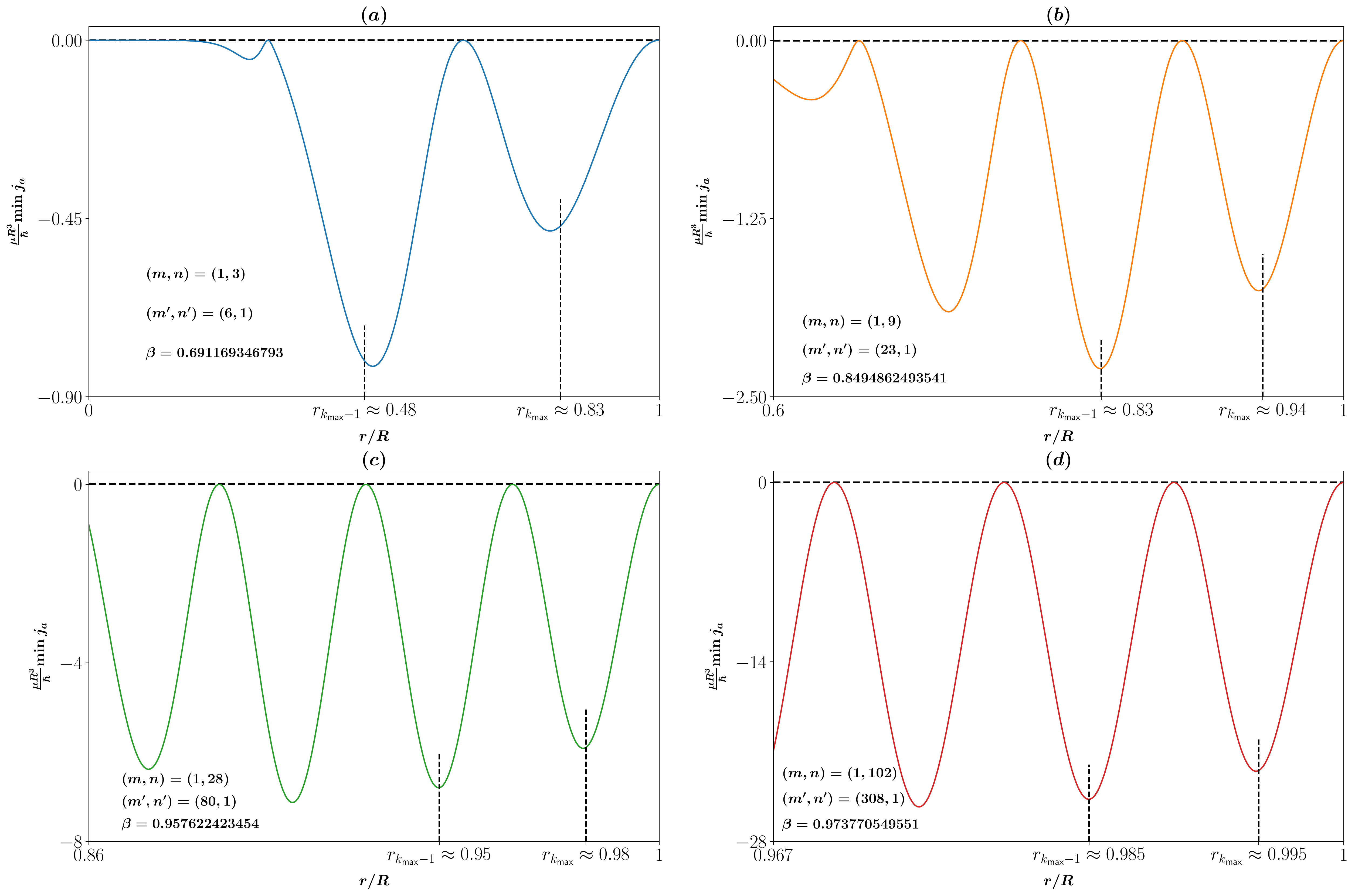}
\caption{(Dimensionless) minimum current $(\mu R^3 / \hbar) \min j_{\text{a}}$, as given by~\eqref{min_j_a_u_v_expr_app}, as a function of the (dimensionless) radial coordinate $r/R$ for the four sets of parameters $(m,n,m',n',\beta)$ considered in Table~\ref{table1}. The corresponding values of the positions $r_{k_{\text{max}}-1}$ and $r_{k_{\text{max}}}$ are also indicated.}
\label{fig_current_with_rk}
\end{figure}
\end{center}

Of course, the integer $k$ in~\eqref{r_k_expr} can actually not be arbitrary, since in view of~\eqref{r_finite} we must have
\begin{equation}
R_1 < r_k = \frac{R}{\gamma_{M n}} \left( k + \frac{M}{2} + \frac{1}{4} \right) \pi < R \, ,
\label{r_k_constraint}
\end{equation}
that is
\begin{equation}
\frac{\gamma_{M n}}{\pi} \frac{R_1}{R} - \frac{M}{2} - \frac{1}{4} < k < \frac{\gamma_{M n}}{\pi} - \frac{M}{2} - \frac{1}{4} \, .
\label{k_range_values}
\end{equation}
Therefore, for completeness we introduce the two integers $k_{\text{min}}$ and $k_{\text{max}}$ defined by
\begin{equation}
k_{\text{min}} \equiv \left\lfloor \frac{\gamma_{M n}}{\pi} \frac{R_1}{R} - \frac{M}{2} - \frac{1}{4} \right\rfloor + 1
\label{k_min_def}
\end{equation}
and
\begin{equation}
k_{\text{max}} \equiv \left\lceil \frac{\gamma_{M n}}{\pi} - \frac{M}{2} - \frac{1}{4} \right\rceil - 1 \, ,
\label{k_max_def}
\end{equation}
where $\lfloor \cdot \rfloor$ and $\lceil \cdot \rceil$ denote the floor and ceiling functions, respectively. The condition $r_k \in (R_1,R)$ hence constrains the integer $k$ in~\eqref{r_k_expr} to take the values
\begin{equation}
k = k_{\text{min}}, \, k_{\text{min}} + 1 , \, \ldots , \, k_{\text{max}} - 1 , \, k_{\text{max}} \, .
\label{k_allowed_values}
\end{equation}
In the sequel, we will focus our attention on the values of $r_k$ that are close to the rim: that is, we will typically consider $k = k_{\text{max}}$. A first advantage of doing this is that, as compared to $k_{\text{min}}$, $k_{\text{max}}$ is independent of the (somehow artificial) parameter $R_1$. An other rationale for considering $k = k_{\text{max}}$ arises from the numerical observation that for values of $k$ close to $k_{\text{max}}$ the particular positions $r_k$ given by~\eqref{r_k_expr} also turn out to accurately describe the locations of the local minima of the minimal current~\eqref{min_j_a_u_v_expr_app} itself. This is indeed illustrated on figure~\ref{fig_current_with_rk} below for the four sets of parameters $(m,n,m',n',\beta)$ considered in Table~\ref{table1}.


\onecolumngrid

\twocolumngrid


Since these particular positions $r_k$ are the (approximate) locations of the extrema of $\phi_{mn}(r)$, they hence precisely correspond to the (approximate) locations of the \textit{maxima} of $\phi_{mn}^2(r)$. In particular, we hence have at the position $r=r_{k_{\text{max}}}$
\begin{equation}
\phi_{mn}^2(r_{k_{\text{max}}}) \sim \frac{1}{R \pi r_{k_{\text{max}}}} \frac{1}{\sin^2 \left( \gamma_{M n} - \frac{M \pi}{2} - \frac{\pi}{4} \right)} \, ,
\label{phi_mn_r_k}
\end{equation}
which is indeed \textit{finite} [the denominator in~\eqref{phi_mn_r_k} being bounded, it can not make the right-hand side in~\eqref{phi_mn_r_k} arbitrarily small]. Of course, this particular position $r_{k_{\text{max}}}$ actually depends on $n$, which itself depends on $M'$ because of the degeneracy condition~\eqref{large_zero}, and thus $r_{k_{\text{max}}}$ also eventually depends on the parameter $u$ that we'll eventually send to infinity. But this is not an issue for what we want to do here: the only important thing for us is that we can actually find such a value of $r$ for any set of parameters $(m,n,m',n',\beta)$ that satisfies the degeneracy condition~\eqref{degeneracy_condition_supp_mat}.


\subsection{The bracketed quantity in\texorpdfstring{~\eqref{min_j_a_ineq}}{}}
\label{brack_qty_subsec}

Now that we found the particular value $r=r_{k_{\text{max}}}$ at which the global factor $\phi_{mn}^2(r)$ in the right-hand side in~\eqref{min_j_a_ineq} reaches a local maximum (and is thus in particular non negligible), we now set $r=r_{k_{\text{max}}}$ into~\eqref{min_j_a_ineq} to get
\begin{align}
\left. \min j_{\text{a}} \right\rvert_{r=r_{k_{\text{max}}}} < \frac{\hbar M}{2 \mu r_{k_{\text{max}}}} \phi_{mn}^2(r_{k_{\text{max}}}) \nonumber \\[0.3cm] \times \left\{ 1 + u v(r_{k_{\text{max}}}) \left[ 1 - \sqrt{1 + \frac{1}{v(r_{k_{\text{max}}})}} \right] \right\} \, .
\label{min_j_a_ineq_r_k}
\end{align}
We now argue that the term $v(r_{k_{\text{max}}}) [ 1 - \sqrt{1 + 1/v(r_{k_{\text{max}}})} ]$ in~\eqref{min_j_a_ineq_r_k} can not take negligible values as $u \to \infty$: this is indeed necessary for our argument, since otherwise the actual value of the bracketed quantity in the right-hand side in~\eqref{min_j_a_ineq_r_k} could not be guaranteed to be arbitrarily negative even in the limit $u \to \infty$.


\begin{table*}[ht]
\centering
\caption{Values of $v(r_{k_{\text{max}}})$ for the four sets of parameters $(m,n,m',n',\beta)$ considered in Table~\ref{table1}.}
\vskip 0.2 cm
\begin{tabular}{ |M{0.7cm}|M{1cm}|M{1cm}|M{0.7cm}|M{2.5cm}|M{3cm}|N }
\hline
\vspace{0.1cm}
$m$ & $n$ & $m'$ & $n'$ & $\beta$ & $v(r_{k_{\text{max}}})$ & \\[0.1cm]
\hline\hline
\vspace{0.1cm}
1 & 3 & 6 & 1 & 0.691169346793 & 1.48137242921629 & \\[0.25cm]
1 & 9 & 23 & 1 & 0.8494862493541 & 1.88363945185356 & \\[0.25cm]
1 & 28 & 80 & 1 & 0.957622423454 & 2.15503028356504 & \\[0.25cm]
1 & 102 & 308 & 1 & 0.973770549551 & 2.32355819487647 & \\[0.1cm]
\hline
\end{tabular}
\label{table2}
\end{table*}


To this end, we propose the following conjecture:
\begin{equation}
v(r_{k_{\text{max}}}) \gtrsim 1 \quad , \quad \text{however large is $u$} \, .
\label{conj}
\end{equation}
Table~\ref{table2} contains numerical evidence of the validity of this conjecture. Similarly to what we did above to numerically investigate the validity of the regime~\eqref{large_u}, here again we restricted our attention to $m=n'=1$: we have then also checked the validity of our conjecture~\eqref{conj} for all the valid sets of values of $(n,m',\beta)$ for $n$ between 3 and 164 (the corresponding valid values of $m'$ being between 6 and 501). Furthermore, for these valid values of $(m,n,m',n',\beta)$, we observed that $v(r_{k_{\text{max}}})$ actually seems to increase with $n$ (and thus also with $m'$): this hence makes our conjecture~\eqref{conj} indeed reasonable.

Now, it is clear numerically that the function $f(v) \equiv v [ 1 - \sqrt{1 + 1/v} ]$ is monotonically decreasing, with the particular values $f(0)=0$, $f(1)=1-\sqrt{2} \approx -0.41$ and $f(\infty)=-1/2$. Therefore, we have in view of our conjecture~\eqref{conj} that $f[v(r_{k_{\text{max}}})] \lesssim f(1) \approx -0.41$, that is more explicitly
\begin{equation}
v(r_{k_{\text{max}}}) \left[ 1 - \sqrt{1 + \frac{1}{v(r_{k_{\text{max}}})}} \right] \lesssim - 0.41 \, ,
\label{finite_factor_v_r_k_max}
\end{equation}
which hence indeed ensures that the term $v(r_{k_{\text{max}}}) [ 1 - \sqrt{1 + 1/v(r_{k_{\text{max}}})} ]$ in the right-hand side of~\eqref{min_j_a_ineq_r_k} can not take negligible values even in the limit $u \to \infty$, limit that we can now safely take.


\subsection{Unboundedness of \texorpdfstring{$\min j_{\text{a}}$}{min j}}
\label{unb_min_j_a_subsec}

In view of~\eqref{finite_factor_v_r_k_max} we readily have (since $u > 0$ by construction)
\begin{equation}
1 + u v(r_{k_{\text{max}}}) \left[ 1 - \sqrt{1 + \frac{1}{v(r_{k_{\text{max}}})}} \right] \lesssim 1 - 0.41 u \, .
\label{bracketed_term}
\end{equation}
Taking the limit $u \to \infty$ in~\eqref{bracketed_term} hence yields
\begin{align}
\lim\limits_{u \to \infty} \left\{ 1 + u v(r_{k_{\text{max}}}) \left[ 1 - \sqrt{1 + \frac{1}{v(r_{k_{\text{max}}})}} \right] \right\} \nonumber \\[0.3cm]
\lesssim \lim\limits_{u \to \infty} \left( 1 - 0.41 u \right) = - \infty \, ,
\label{lim_u_inf_ineq}
\end{align}
so that the left-hand side itself must tend to $-\infty$, i.e.
\begin{equation}
\lim\limits_{u \to \infty} \left\{ 1 + u v(r_{k_{\text{max}}}) \left[ 1 - \sqrt{1 + \frac{1}{v(r_{k_{\text{max}}})}} \right] \right\} = - \infty \, .
\label{lim_u_inf}
\end{equation}

Therefore, we now take the limit $u \to \infty$ in~\eqref{min_j_a_ineq_r_k}: we get
\begin{align}
\lim\limits_{u \to \infty} \left. \min j_{\text{a}} \right\rvert_{r=r_{k_{\text{max}}}} < \lim\limits_{u \to \infty} \frac{\hbar M}{2 \mu r_{k_{\text{max}}}} \phi_{mn}^2(r_{k_{\text{max}}}) \nonumber \\[0.3cm]
\times \left\{ 1 + u v(r_{k_{\text{max}}}) \left[ 1 - \sqrt{1 + \frac{1}{v(r_{k_{\text{max}}})}} \right] \right\} \, ,
\label{min_j_a_lim_ineq}
\end{align}
that is in view of~\eqref{lim_u_inf}, and precisely because we constructed $r_{k_{\text{max}}}$ so as to ensure that the prefactor $(\hbar M/2 \mu r_{k_{\text{max}}}) \phi_{mn}^2(r_{k_{\text{max}}})$ in~\eqref{min_j_a_lim_ineq} remains \textit{finite},
\begin{equation}
\lim\limits_{u \to \infty} \left. \min j_{\text{a}} \right\rvert_{r=r_{k_{\text{max}}}} < - \infty \, ,
\label{min_j_a_lim_ineq_infty}
\end{equation}
that is finally
\begin{equation}
\lim\limits_{u \to \infty} \left. \min j_{\text{a}} \right\rvert_{r=r_{k_{\text{max}}}} = - \infty \, .
\label{min_j_a_unbounded}
\end{equation}
This hence demonstrates that the spatially local current $j_{\text{a}}$ is indeed unbounded from below.



\bibliography{DATABASE_Quantum_Backflow}



\end{document}